\documentstyle[prl,aps]{revtex}

\begin{document}

\draft

\title{ A New Stochastic Interpretation of Quantum Mechanics}

\author{Robert Alicki}

\address{Institute of Theoretical Physics and Astrophysics, University
of Gda\'nsk, Wita Stwosza 57, PL 80-952 Gda\'nsk, Poland}

\date{\today}
\maketitle

\begin{abstract}
The reinterpretation of quantum mechanical formalism in terms of
a classical model with a continuous material "$\Psi$ -field" acting upon
a point-like particle which is subjected to large friction and random
forces is proposed. This model gives a mechanism for sudden
"quantum jumps" 
and provides a simple explanation of "Schr\"odinger Cat" phenomena.
\end{abstract}


Although the formalism of quantum mechanics and its predictions
within the generally accepted limits of applicability are beyond any
doubt the interpretation of the fundamental object-- the wave function
$\Psi$-- is still a topic of debate at least for a small but dedicated
group of physicist and philosophers\cite{[1]}. The standard Copenhagen
interpretation which states that $|\Psi (X,t)|^2$ is the probability
density of finding at the time $t$ the physical system in a point $X$
of the configurations space produces an elegant mathematical formalism
of Quantum Probability and a consistent theory with an enormous 
predictive power. Nevertheless the immaterial  nature 
of the continuous wave function
contrasted with a point-like structure of elementary  particle 
reconciliated finally within the Bohr's "wave-particle dualism"
disturbs from time to time a good frame of mind of not only undergraduate
students but also professional physicists.
Therefore the critical discussion of alternative interpretations is
not completely pointless at least from the pedagogical point of view.
Moreover there exists a class of phenomena which are difficult to reconcile
with the orthodox first-principles quantum mechanics. One can mention
in this context the following topics: the "problem of molecular structure"
\cite{[2]} in particular the existence of optical isomers, 
Schr\"odinger Cat paradox or more generally the absence of 
superpositions of macroscopically distinguishable states. 
The interpretation of quantum theory proposed in this note provides
a simple explanation of these phenomena preserving at the same time
the mathematical formalism of quantum mechanics as a basic ingredient. 
The main idea may look similarly
to the ideas of Nelson \cite{[3]} 
and  Bohm interpretation of quantum mechanics \cite{[4]} 
but in fact is quite different and essentially simpler. 
In contrast to Nelson stochastic mechanics which is 
mathematically equivalent to the standard quantum theory the presented 
formalism reproduces quantum mechanical predictions 
under certain additional conditions. Unlike in classical field theories 
(e.g. electrodynamics, gravitation theory) the fundamental object 
is not a space-time but a configuration space $\Omega$ of
the physical system (possible the whole Universe) and an absolute 
(cosmological) time $t$ as a parameter. Two different entities live 
on $\Omega$: the wave function $\Psi (X,t), X\in \Omega$ and a point-like 
"Particle" described by the trajectory $X(t)\in\Omega$. Here $\Psi$
is a really existing "classical" field which can be detected by its influence
on the Particle. The time evolution is given by a pair of
equations: the Schr\"odinger equation for $\Psi$
$${\partial\over {\partial t}}\Psi(X,t) = -{i\over{\hbar}}{\cal H}
\Psi(X,t) \eqno(1)$$
where ${\cal H}$ is the Hamiltonian operator which contains all information
about the structure of space-time and symmetries of the theory
and the equation of motion describing the trajectory $X(t)$. The later
depends on the structure of $\Omega$ and we restrict our discussion to the 
simplest case of $\Omega ={\bf R}^N$. In this case one can propose 
the following Langevin equation describing a Particle exercising
a Brownian motion on $\Omega$ under the influence of an external time 
dependent potential $V(X,t) = -\ln |\Psi (X,t)|^2$
$$ {d\over {dt}}X(t) = \lambda {\nabla}_X \ln |\Psi(X(t),t)|^2 +
\sqrt{2\lambda} \xi(t)\eqno(2)$$
where $\xi (t) = \{\xi_1(t),...,\xi_N (t)\}$ is a universal white noise
satisfying
$$ <\xi_k(t) \xi_l(s)> = \delta_{kl}\delta (t-s)\eqno(3)$$
and  with the diffusion constant $\lambda$ which will 
take large enough value. The logarithmic form of the potential allows
to separate degrees of freedom when $\Psi$ possesses a product structure.
A possible value of $\lambda$ can be parameterized as follows
$$ \lambda = {{l_{\Omega}^2}\over {\tau}} \eqno(4)$$
where $l_{\Omega}$ is a characteristic "length scale" on the configuration
space (e.g. Bohr radius in the atomic physics) and $\tau$ is a very short
universal time scale of diffusion  (e.g. Planck unit of time).
This is a fundamental difference in comparison with other
"stochastic approaches" where the diffusion constant is 
proportional to $\hbar$ and hence "small". 
The rapid diffusion in our model describes in classical terms the so-called 
"quantum jumps"\cite{[5]}. The possible mechanism producing the friction 
and random forces in Eq.(2) is not specified here
but it can be related either to a kind of "vacuum fluctuations" or due to 
the interaction between the enormous number of degrees of freedom.
The probability distribution of the Particle position $p(X,t)$
satisfies the Smoluchowski equation completely equivalent to Eq.(2)
$${{\partial p}\over {\partial t}} = \lambda \nabla_X [-\nabla_X
(\ln |\Psi|^2)p + \nabla_X p]\ .\eqno(5)$$
For a very large $\lambda $ and if the potential 
$V(X,t) = -\ln |\Psi (X,t)|^2$
does not produce infinite or high practically impenetrable barriers between 
different regions
of the configurations space $\Omega$ the probability distribution 
$p(X,t)$ relaxes rapidly to its temporal equilibrium
value such that the following adiabatic approximation is valid
$$ p(X,t) = Z^{-1} |\Psi(X,t)|^2 + o(\lambda ^{-1})\eqno(6)$$
where $Z$ is a normalizing constant ($|\Psi|^2$ need not to be normalized
to 1). Hence up to the short time relaxation effects which can be 
unobservable under the conditions of above on $|\Psi|^2$ and $\lambda $ 
all statistical  predictions of
quantum mechanical formalism concerning the functions of $X$ are recovered. 
It seems that the information about others observables (e.g.
momentum) are obtained analyzing the time evolution of wave
packets in configuration space. Hence in this case all measurable properties
of the physical system are determined by the wave equation (1).

The basic assumption concerning the absence of impenetrable barriers
of the potential $-\ln |\Psi (X,t)|^2$ is reasonable for the microscopic 
systems for which the superpositions of different quickly oscillating
eigenstates of the Hamiltonian smear out the nodes of $|\Psi|^2$ and allows
the rapid relaxation of $p(X,t)$ to the value proportional to $|\Psi(X,t)|^2$.
However, in many cases the above picture is false. In a standard interference
experiment with a single particle the nodes of the wave function restrict
the random trajectory of the particle to a vinicity of the local maximum of
$\Psi$. Therefore the actual trajectory of the particle is rather localized
and depends sensitively on the initial conditions. Nevertheless repeating
the experiment with a large number of particles we recover again the
interference pattern predicted by the standard quantum theory. Like for the
standard theory any interaction with a measuring device introduces extra
degrees of freedom to the configuration space and destroys the interference
as expected.

A different situation is illustrated by the following simple
example.
Consider a physical system of a single degree of freedom described by the
wave function which is a superposition of two Gaussians of the width $a$
and separated by the distance $2b$. Then
$$\Psi (x) = \exp \{(x-b)^2/2a^2\} + \exp \{(x+b)^2/2a^2\} \ .\eqno(7)$$
The relaxation time for the corresponding double-well potential 
$-\ln |\Psi|^2$ can be roughly estimated by the Kramers formula\cite{[6]}
for the escape time $T$ from the well which in our case is given by
$(b>>a)$
$$ T \approx {a^3\over {\lambda b}}\exp(b^2/a^2) \ .\eqno(8)$$
The rapid increase of the relaxation time with the ratio $b/a$ implies
(for the large $b/a$) the localization of the probability distribution 
$p(x,t)$ at a given well. The similar arguments provide
the plausible explanation  of the observed localization
effects in optical isomers or macroscopic systems.
\par
One can imagine a few possible modifications of the presented approach. 
The configuration space $\Omega$ can be replaced by the phase-space $\Gamma$
if one uses the phase-space formulation of the Eq.(1) and an obvious extension
of the Eqs.(2,3,5) to $\Gamma$. The linearity of the Schr\"odinger equation
is not a fundamental principle anymore and small nonlinear perturbations
in particular describing the influence of the Particle on $\Psi$ are 
conceivable.
\par
In conclusion we summarize several appealing features of the discussed approach.
\par
1) The proposal is rather conservative: the most important object to be
calculated first is the quantum state $\Psi(X,t)$ interpreted here as 
a classical field on $\Omega$.
\par
2) The formalism is not equivalent to the quantum mechanics, in particular
there should be possible to test experimentally the failures of the 
adiabatic approximation (see Eq.(6)).
\par
3) The underlying diffusion process with a large diffusion constant provides
a physical mechanism of sudden "quantum jumps" between different states
of the system.
\par
4) The absence of superpositions of macroscopically distinguishable states
is easy to explain. 
\par
5) As the diffusion equation (4) is irreversible the time arrow is build
into the formalism.
\par
6) The inconceivable world of quanta is replaced again by the classical
notions of fields and point particles living, however, on the 
multidimensional configuration space.

\acknowledgments
The author thanks R. Horodecki for the comments on the manuscript.
The work is supported by the Grant KBN BW 5400-5-0303-7.

\end{document}